\documentclass[twocolumn,prd,aps,amsmath,amssymb,nofootinbib]{revtex4-1}
\usepackage{graphics}
\usepackage{graphicx}
\usepackage{dcolumn}
\usepackage{bm}
\usepackage{mathrsfs}
\usepackage{pstricks}
\usepackage{color}
\usepackage{soul}
\usepackage{ulem}
\usepackage{float}
\usepackage{dsfont}
\usepackage{slashed}
\usepackage{amsmath}
\usepackage{epsfig}
\usepackage{amsfonts}
\usepackage{amssymb}
\usepackage{comment}
\def\beq{\begin{equation}}
\def\eeq{\end{equation}}
\def\bea{\begin{eqnarray}}
\def\eea{\end{eqnarray}}

\newcommand{\vect}[1]{\boldsymbol{#1}}

\begin{document}

\title{New BCH-like relations of the \textit{su}$(1,1)$, \textit{su}$(2)$ and \textit{so}$(2,1)$ Lie algebras}
\author{D. M. Tibaduiza}
\email{Correspondence to: danielmartinezt@gmail.com}
\affiliation{Instituto de F\'isica, Universidade Federal Fluminense, \\ 
Niteroi, RJ 24210-346, Brazil}
\author{A.~H.~Arag\~{a}o}
\affiliation{Instituto de F\'isica, Universidade Federal de Goi\'as, \\ 
Goi\^{a}nia, GO 74690-900, Brazil}
\author{C. Farina}
\author{C.~A.~D.~Zarro}
\affiliation{Instituto de F\'isica, Universidade Federal do Rio de Janeiro, \\ 
Rio de Janeiro, RJ 21941-972, Brazil}


\begin{abstract}

In this work we demonstrate new BCH-like relations involving the generators of the 
\textit{su}$(1,1)$, \textit{su}$(2)$ and \textit{so}$(2,1)$ Lie algebras. 
We use our results to obtain in a straightforward way the composition of an arbitrary number of elements of the corresponding Lie groups. 
In order to make a self-consistent check of our results, as a first application we recover the non-trivial composition law of two arbitrary squeezing operators. 
As a second application, we show how our results can be used to compute the time evolution operator of physical systems 
described by time-dependent hamiltonians given by linear combinations of the generators of the aforementioned Lie algebras.
\end{abstract}



\maketitle

\section{Introduction}\label{intro}

Symmetries are a cornerstone concept in physics and group theory is the natural language to investigate them.
Among all groups, Lie groups are of great interest in physics since they are, grossly speaking, continuous groups with the structure 
of a differentiable manifold. This structure enables one to introduce special vectors called infinitesimal 
generators which span the tangent space near the identity element. Then, defining the 
commutation operation between such generators, one can define the so-called Lie algebras. 
A natural question that arises is how to compose an arbitrary number of elements of a Lie group.
This question was investigated by J. Campbell, H. Baker and F. Hausdorff and their results 
as well as similar ones are referred to as BCH-like relations 
\cite{Campbell-1896,Campbell-1897,Baker-1901,Baker-1905,Hausdorff-1906}. 
For an interested reader, we refer to Ref. \cite{SZEKERES-2004} for an introduction 
from a geometric point of view, and Ref. \cite{GILMORE-2012} for a more traditional presentation of these issues.

Since symmetries and degeneracies of physical systems, particularly degeneracies of atomic spectra, 
are closely related, the first physical applications of Lie groups and Lie algebras in quantum mechanics 
came from this kind of study. 
The algebraic approach consists, basically, in writing a set of quantum mechanical 
operators, that commute with the hamiltonian of the system, as elements of a Lie algebra 
describing the symmetries of such system.  
A well known example is the rotational symmetry described by an $SO(3)$ group, exhibited by all 
physical systems where the interaction is governed by central potentials. 
This symmetry is responsible for the $2\ell + 1$ degenerate states with the same 
angular momentum quantum number $\ell$ but different quantum numbers $m$ 
(related to the eigenvalues of operator $\hat L_z$). 
The hydrogen atom, for instance, exhibits an additional degeneracy, since the corresponding 
energy levels do not depend on the quantum number $\ell$, so that associated to a given energy level, 
$E_n$, there are $\sum_{\ell = 0}^{n-1} (2\ell + 1) = n^2$ degenerate states. 
This extra degeneracy was unveiled in a seminal paper by W. Pauli in 1926  \cite{Pauli-1926} 
in which he used a purely algebraic solution to calculate the hydrogen energy spectrum. 
Pauli's  strategy was to construct, from the angular momentum and the Laplace-Runge-Lenz vector operators, 
the six generators of the $so(4)$ Lie algebra. For a detailed discussion of Pauli's method see 
Ref. \cite{Borowitz-Book-1967} (see also Ref. \cite{LANDAU-1977}). 
For completeness, it is worth mentioning that the continuous spectra of the 
hydrogen atom has the Lorentz algebra $so(3,1)$ as the Lie algebra describing its symmetries, 
which can be used in the computation of the corresponding phase shifts for the Coulomb potential \cite{ZWANZIGER-1967}. 

Methods involving Lie groups and Lie algebras can also be used to deal with dynamical properties of physical systems, such as the computation of time evolution operators, Feynman propagators and Green functions of non-trivial systems (see, $e.g.$ Refs. \cite{Vaidya:1989tj, BoschiFilho:1990az} and references therein). In this sense, they are very important in solving, not only the spectrum of a hamiltonian, but also to obtain, in a fairly elegant way, the time evolution of quantum states of certain physical systems. 
Since algebraic methods  are not restricted to deal only with time-independent hamiltonians, these methods may provide alternative routes for solving problems described by time-dependent hamiltonians without the necessity of using Schwinger-Dyson formalism.

Algebraic methods based on the Lie algebras $so(2,1)$, $su(2)$ and $su(1,1)$, can be applied in many physical systems. 
The algebra $so(2,1)$ has, for instance, manifold applications: three-dimensional harmonic oscillator, Morse oscillator \cite{CORDERO-1970}, Benzene molecule which is described by the Hartmann potential \cite{Vaidya:1989tj, GERRY-1986} and polyatomic molecules modeled by a P\"{o}sch-Teller potential \cite{BARUT-1987}, among others. In all of these systems, the propagator as well as the spectrum can be obtained by an algebraic method based on the $so(2,1)$ Lie algebra. 
As the group $SO(3)$ has an universal covering group $SU(2)$, the Lie algebra $su(2)$ 
can be used to investigate the angular momentum algebra of systems that exhibit rotational symmetry. 
The Lie algebra $su(1,1)$ can be applied, for instance, to quantum optics. Two relevant examples are 
coherent \cite{BARUT-1971} and squeezed states \cite{Truax-1985, Rhodes-1989} 
of the quantized electromagnetic field, both frequently described by using algebraic methods based on the $su(1,1)$ Lie algebra. 
A natural scenario to study these states of light is the driven time-dependent harmonic oscillator.
In Ref. \cite{DMT-JPA-2020} it was developed a method, based also on the $su(1,1)$ Lie algebra, 
for solving a time-dependent harmonic oscillator for any initial state and with an arbitrary time-dependent frequency. 
Particularly, this method is very convenient for numerical calculations, since the final expressions are written as general continued fractions. 
A relatively simple but non-trivial and very instructive example of a harmonic oscillator with a time-dependent frequency, 
which admits analytical solution by algebraic methods, can be found in Ref. \cite{DMT-BJP-2020}. 
For a comprehensive reference on physical applications of $su(2)$ and $su(1,1)$ Lie algebras, we suggest Ref. \cite{INOMATA-1992}.

In this work we establish, simultaneously, new BCH-like relations for the 
\textit{su}$(1,1)$, \textit{su}$(2)$ and \textit{so}$(2,1)$ Lie algebras. 
Our results arise from the composition of two elements of the correspondent Lie groups and 
we initially use them to establish a novel iterative analytical form to calculate 
the composition rule of an arbitrary number of elements of the corresponding Lie groups. 
We show that this general rule can be conveniently written in terms of a general continued fraction, 
which has a very useful structure to be implemented computationally. 
As a first application of our results, and in order to check the validity of our results, 
we reobtain, in a novel way, the non-trivial composition law of two arbitrary squeezing operators. 
We show explicitly that the resulting operator of such composition is not necessarily another squeezing operator, 
but the product of a squeezing operator and a rotation operator. 
As a second application we show how our results can be used to compute the time evolution 
operator of any physical system described by a time-dependent hamiltonian 
expressed by linear combinations of the generators of the aforementioned Lie algebras.


\section{Simultaneous treatment of the three Lie algebras}\label{RF}

Consider the following commutation relations
\begin{equation}
\left[\hat{T}_{-},\hat{T}_{+}\right]=2\epsilon\hat{T}_{c} \:\: \mbox{and} \:\: \left[\hat{T}_{c},\hat{T}_{\pm}\right]=\pm\delta\hat{T}_{\pm}\, ,
\label{eq:algebraK}
\end{equation}
where the parameters $\epsilon$ and $\delta$ are introduced in order to treat the mentioned Lie algebras simultaneously. In fact, note that 
$\hat{T}_{+}$, $\hat{T}_{c}$ and $\hat{T}_{-}$ can be identified as the three generators of the 
\textit{su}(1,1), \textit{su}$(2)$ and \textit{so}$(2,1)$ Lie algebras \cite{GILMORE-2012} 
once we choose $(\epsilon,\delta)=(1,1)$, $(\epsilon,\delta)=(-1,1)$ and $(\epsilon,\delta)=(\frac{i}{2},i)$, respectively. 
Let us define the generic operator 
\begin{equation}
\hat{G}=\hat{f}(\vect{\lambda})=e^{\lambda_{+}\hat{T}_{+}+\lambda_{c}\hat{T}_{c}+\lambda_{-}\hat{T}_{-}} \, ,
\label{eq:Gj}
\end{equation}
where $\vect{\lambda}=(\lambda_{+},\lambda_{c},\lambda_{-})$ is a set of arbitrary complex parameters. 
Operator $\hat{G}$ is, therefore, an arbitrary element of any of the correspondent Lie groups. Since any element of a Lie group  
can also be represented as a product of exponentials of the corresponding Lie algebra generators \cite{GILMORE-2012}, 
the same $\hat{G}$ can be written in the following suitable representation
\begin{equation}
\hat{G}=\hat{h}(\vect{\Lambda}) = e^{\Lambda_{+}\hat{T}_{+}}e^{\ln(\Lambda_{c})\hat{T}_{c}}e^{\Lambda_{-}\hat{T}_{-}}  \, ,
\label{eq:BCH1}
\end{equation}
where $\vect{\Lambda}=(\Lambda_{+},\Lambda_{c},\Lambda_{-})$ is a new set of parameters. 
The relations connecting these new parameters  with the old ones, $(\lambda_{+},\lambda_{c},\lambda_{-})$,  can be obtained 
by imposing that $\hat{f}(\vect{\Lambda})=\hat{h}(\vect{\lambda})$. It can be shown that this procedure leads to the following relations\footnote{Explicit demonstrations of these relations for the \textit{su}(1,1) and \textit{su}$(2)$ Lie algebras can be found 
in Refs. \cite{Truax-1985, DMT-BJP-2020}. Alternatively, these relations can also be deduced by using the unitary expansion method of Ref. \cite{Zagury-2010}.}
\begin{eqnarray}
\label{truej4}
\Lambda_{c}&=&\left(\cosh(\nu)-\frac{\delta\lambda_{c}}{2\nu} \sinh(\nu)\right)^{-\frac{2}{\delta}},\\
\label{truej5}
\Lambda_{\pm}&=&\frac{2\lambda_{\pm} \sinh(\nu)}{2\nu \cosh(\nu)-\delta\lambda_{c}\sinh(\nu)} \, ,
\end{eqnarray}
with $\nu$ given by
\begin{equation}
\label{truej6}
\nu^{2} = \left(\frac{\delta\lambda_{c}}{2}\right)^{2}-\delta\epsilon\lambda_{+}\lambda_{-} \, .
\end{equation}
The above relations are known as BCH-like relations of the given Lie algebras.



\section{New BCH-like relations and proof of the composition rule}\label{DEMO} 

We aim to write
\begin{align}
&\hat{G}(\vect{\Lambda}_{N})\hat{G}(\vect{\Lambda}_{N-1})\cdots\hat{G}(\vect{\Lambda}_{2})\hat{G}(\vect{\Lambda}_{1})= \nonumber\\[4pt]
&=\hat{G}(\Lambda_{N+},\Lambda_{Nc},\Lambda_{N-})\hat{G}(\Lambda_{(N-1)+},\Lambda_{(N-1)c},\Lambda_{(N-1)-})\cdots \nonumber\\[4pt]
&\cdots \hat{G}(\Lambda_{2+},\Lambda_{2c},\Lambda_{2-})\hat{G}(\Lambda_{1+},\Lambda_{1c},\Lambda_{1-}) \, ,
\end{align}
as a unique group element given by 
%
$\hat{G}(\alpha_{N},\beta_{N},\gamma_{N})$. 
%
Our purpose is to 
write parameters $(\alpha_{N}, \beta_{N}, \gamma_{N})$ 
in terms of coefficients $(\Lambda_{i+}, \Lambda_{ic}, \Lambda_{i-})$ with $(i=1,2,\cdots,N)$. 
With this goal in mind, we start by calculating the composition of just two of them, namely, 
\begin{eqnarray}
\hat{G}(\vect{\Lambda}_{2})\hat{G}(\vect{\Lambda}_{1})&=&
e^{\Lambda_{2+}\hat{T}_{+}}e^{\ln(\Lambda_{2c})\hat{T}_{c}}e^{\Lambda_{2-}\hat{T}_{-}}\cdot \cr\cr
&& \cdot e^{\Lambda_{1+}\hat{T}_{+}}e^{\ln(\Lambda_{1c})\hat{T}_{c}}e^{\Lambda_{1-}\hat{T}_{-}}  \, ,
\label{eq:2compo}
\end{eqnarray}
where the symbol $\cdot$ reinforces group multiplication whenever a line changes. 
The method we shall use consists in reordering the above product of exponentials until obtaining the same structure as in Eq. (\ref{eq:BCH1}). 
Initially we want to move all operators associated to $\hat{T}_{+}$ to the left (or, equivalently, all operators associated to $\hat{T}_{-}$ to the right). 
Inserting in the previous equation the identity as $\mathds{1}=e^{\Lambda_{1+}\hat{T}_{+}}e^{-\Lambda_{1+}\hat{T}_{+}}$, 
we can write it as
\begin{eqnarray}
\hat{G}(\vect{\Lambda}_{2})\hat{G}(\vect{\Lambda}_{1})&=&e^{\Lambda_{2+}\hat{T}_{+}}e^{\ln(\Lambda_{2c})\hat{T}_{c}} e^{\Lambda_{1+}\hat{T}_{+}}\cdot \cr\cr
&&\cdot \left\{e^{-\Lambda_{1+}\hat{T}_{+}}e^{\Lambda_{2-}\hat{T}_{-}}e^{\Lambda_{1+}\hat{T}_{+}}\right\} \cdot \cr\cr
&& \cdot e^{\ln(\Lambda_{1c})\hat{T}_{c}}e^{\Lambda_{1-}\hat{T}_{-}}  \, .
\label{eq:2compomod1bis}
\end{eqnarray}
Using the following BCH relations, valid for generic operators $\hat{A}$, $\hat{B}$ and $\hat{C}$ \cite{Barnett-Book-1997},
\begin{eqnarray}
\label{eq:BCHclasic}
e^{\hat{A}}\hat{B}e^{-\hat{A}}&=&\hat{B}+\left[\hat{A},\hat{B}\right]+\frac{1}{2!}\left[\hat{A},\left[\hat{A},\hat{B}\right]\right]+
\ldots \, 
\end{eqnarray}
and
\begin{equation}
e^{\hat{A}}f\left(\hat{C}\right)e^{-\hat{A}}=f\left(e^{\hat{A}}\hat{C}e^{\hat{-A}}\right)\, ,
\label{eq:BCHfunc}
\end{equation}
%
as well as the commutation rules written in Eq. (\ref{eq:algebraK}), the term inside the brackets of Eq. (\ref{eq:2compomod1bis}) can be written as 
\begin{eqnarray}
e^{-\Lambda_{1+}\hat{T}_{+}}e^{\Lambda_{2-}\hat{T}_{-}}e^{\Lambda_{1+}\hat{T}_{+}}&=&e^{\Lambda_{2-} \left\{e^{-\Lambda_{1+}\hat{T}_{+}}\hat{T}_{-}e^{\Lambda_{1+}\hat{T}_{+}}\right\}} \cr\cr
&=& e^{\Lambda_{2-} \left\{\epsilon\delta(\Lambda_{1+})^{2}\hat{T}_{+}+2\epsilon\Lambda_{1+}\hat{T}_{c}+\hat{T}_{-}\right\}} \cr\cr
&=& e^{\sigma_{+}\hat{T}_{+}+\sigma_{c}\hat{T}_{c}+\sigma_{-}\hat{T}_{-}} \, ,
\label{eq:embr1bch}
\end{eqnarray}
where 
\begin{equation}
\sigma_{+}=\epsilon\delta\Lambda_{2-}(\Lambda_{1+})^{2}  \: \mbox{,} \:\:\:\:\: \sigma_{c}=2\epsilon\Lambda_{2-}\Lambda_{1+} \:\:\: \mbox{and} \:\:\: 
\sigma_{-}=\Lambda_{2-} \, .\nonumber 
\label{eq:sigmas}
\end{equation}
Using Eqs. (\ref{eq:Gj})--(\ref{truej6}) on the right side of Eq. (\ref{eq:embr1bch}), we obtain
\begin{equation}
e^{-\Lambda_{1+}\hat{T}_{+}}e^{\Lambda_{2-}\hat{T}_{-}}e^{\Lambda_{1+}\hat{T}_{+}}=e^{\Sigma_{+}\hat{T}_{+}}e^{\ln(\Sigma_{c})\hat{T}_{c}}e^{\Sigma_{-}\hat{T}_{-}},
\label{eq:embr1bchfin}
\end{equation}
with
\begin{eqnarray}
\label{eq:Sigmapfin}
&&\Sigma_{+}=\frac{\epsilon\delta(\Lambda_{1+})^{2}\Lambda_{2-}}{1-\epsilon\delta\Lambda_{2-}\Lambda_{1+}} \, , \:\:\:\:
\Sigma_{c}=\left(1-\epsilon\delta\Lambda_{2-}\Lambda_{1+}\right)^{-\frac{2}{\delta}} \,   \cr
&&\:\:\:\:\:\:\:\:\:\:\:\:\:\:\:\:\:\:\:\:\:\mbox{and} \:\:\: \Sigma_{-}=\frac{\Lambda_{2-}}{1-\epsilon\delta\Lambda_{2-}\Lambda_{1+}} \, .
\end{eqnarray}
Substitution of Eq. (\ref{eq:embr1bchfin}) into Eq. (\ref{eq:2compomod1bis}) results in
\begin{align}
\hat{G}(\vect{\Lambda}_{2})\hat{G}(\vect{\Lambda}_{1})\,\,\,=&\,\,\,e^{\Lambda_{2+}\hat{T}_{+}}\left\{e^{\ln(\Lambda_{2c})\hat{T}_{c}} 
e^{(\Lambda_{1+}+\Sigma_{+})\hat{T}_{+}}\right\}\cdot \nonumber\\[4pt]
&\cdot e^{\ln(\Sigma_{c})\hat{T}_{c}}e^{\Sigma_{-}\hat{T}_{-}} e^{\ln(\Lambda_{1c})\hat{T}_{c}}e^{\Lambda_{1-}\hat{T}_{-}}.
\label{eq:2compomod2}
\end{align}
Now, we want to reorder the embraced product in the last equation, such that the $\hat{T}_{+}$ operator goes to the left. Following a similar procedure as before, we obtain
\begin{align}
e^{\ln(\Lambda_{2c})\hat{T}_{c}}&e^{(\Lambda_{1+}+\Sigma_{+})\hat{T}_{+}} = \nonumber\\[4pt]
&=\left\{e^{\ln(\Lambda_{2c})\hat{T}_{c}}e^{(\Lambda_{1+}+\Sigma_{+})\hat{T}_{+}}e^{-\ln(\Lambda_{2c})\hat{T}_{c}}\right\}e^{\ln(\Lambda_{2c})\hat{T}_{c}} \nonumber\\[5pt]
&=e^{\left\{(\Lambda_{1+}+\Sigma_{+})(\Lambda_{2c})^{\delta}\hat{T}_{+}\right\}}e^{\ln(\Lambda_{2c})\hat{T}_{c}}.
\label{eq:embr3}
\end{align}
Substituting the above result into Eq. (\ref{eq:2compomod2}), we have
\begin{eqnarray}
\hat{G}(\vect{\Lambda}_{2})\hat{G}(\vect{\Lambda}_{1})&=&
e^{\bigl(\Lambda_{2+}+(\Lambda_{1+}+\Sigma_{+})(\Lambda_{2c})^{\delta}\bigl)\hat{T}_{+}}e^{\ln(\Lambda_{2c}\Sigma_{c})\hat{T}_{c}} \cdot \cr\cr
&&\cdot \left\{e^{\Sigma_{-}\hat{T}_{-}}e^{\ln(\Lambda_{1c})\hat{T}_{c}}\right\}e^{\Lambda_{1-}\hat{T}_{-}}  \, ,
\label{eq:2compomod3}
\end{eqnarray}
where we used that 
%
$e^{\ln(\Lambda_{2c})\hat{T}_{c}} e^{\ln(\Sigma_{c})\hat{T}_{c}} = 
e^{\ln(\Lambda_{2c}\Sigma_{c})\hat{T}_{c}} \, .$
%
In order to cast all operators associated to $\hat{T}_{c}$ at the center of the expression, the embraced expression is rewritten by the same token as:
\begin{align}
e^{\Sigma_{-}\hat{T}_{-}}e^{\ln(\Lambda_{1c})\hat{T}_{c}}\,\,=&\,\, 
e^{\ln(\Lambda_{1c})\hat{T}_{c}}\left\{e^{-\ln(\Lambda_{1c})\hat{T}_{c}}e^{\Sigma_{-}\hat{T}_{-}}e^{\ln(\Lambda_{1c})\hat{T}_{c}}\right\} \nonumber\\[4pt]
=&\,\, e^{\ln(\Lambda_{1c})\hat{T}_{c}}e^{\left\{\Sigma_{-}(\Lambda_{1c})^{\delta}\hat{T}_{-}\right\}} \, .
\label{eq:embr4}
\end{align}
Substitution of the above result in Eq. (\ref{eq:2compomod3}) yields
\begin{align}
\hat{G}(\vect{\Lambda}_{2})\hat{G}(\vect{\Lambda}_{1})\,\,=&\,\, e^{\bigl(\Lambda_{2+}+(\Lambda_{1+}+\Sigma_{+})(\Lambda_{2c})^{\delta}\bigl)\hat{T}_{+}}
e^{\ln(\Lambda_{1c}\Lambda_{2c}\Sigma_{c})\hat{T}_{c}} \cdot \nonumber\\[5pt]
&\,\, \cdot e^{(\Sigma_{-}(\Lambda_{1c})^{\delta}+\Lambda_{1-})\hat{T}_{-}} \, .
\label{eq:befSig}
\end{align}
Finally, inserting the values of $\Sigma_{+}$, $\Sigma_{-}$ and $\Sigma_{c}$, given by Eq. (\ref{eq:Sigmapfin}), into the above equation, we find
\begin{equation}
\hat{G}(\vect{\Lambda}_{2})\hat{G}(\vect{\Lambda}_{1})=e^{\alpha_{2}\hat{T}_{+}}
e^{\ln(\beta_{2})\hat{T}_{c}}e^{\gamma_{2}\hat{T}_{-}} \, ,
\label{eq:2compgamalta}
\end{equation}
where we define
\begin{align}
\label{beta2}
&\alpha_{2}=\Lambda_{2+}+\frac{\Lambda_{1+}(\Lambda_{2c})^{\delta}}{1-\epsilon\delta\Lambda_{1+}\Lambda_{2-}} \, , \:\:\:\:\:\:
\beta_{2}=\frac{\Lambda_{1c}\Lambda_{2c}}{\left(1-\epsilon\delta\Lambda_{1+}\Lambda_{2-}\right)^{\frac{2}{\delta}}} \,   \nonumber\\[4pt]
&\:\:\:\:\:\:\:\:\:\:\:\:\:\:\:\:\:\:\:\:\:\mbox{and} \:\:\: \gamma_{2}=\Lambda_{1-}+\frac{\Lambda_{2-}(\Lambda_{1c})^{\delta}}{1-\epsilon\delta\Lambda_{1+}\Lambda_{2-}} \, .
\end{align}
Then, Eq. (\ref{eq:2compgamalta}) is clearly another $\hat{G}$ operator. 
Equations (\ref{eq:2compgamalta}) and (\ref{beta2}) are the new BCH-like relations.

Now, we compose three group elements. Hence, let us compose the product of 
Eq. (\ref{eq:2compgamalta}) with a third group element, namely,
\begin{equation}
\hat{G}(\vect{\Lambda}_{3})\hat{G}(\vect{\Lambda}_{2})\hat{G}(\vect{\Lambda}_{1})=
\hat{G}(\Lambda_{3+},\Lambda_{3c},\Lambda_{3-})\hat{G}(\alpha_{2},\beta_{2},\gamma_{2}) \, . \\
\label{eq:3compgamalta}
\end{equation}
The right hand side of the above equation is the composition of just two elements. 
Then, using the new BCH-like relations, Eqs. (\ref{eq:2compgamalta}) and (\ref{beta2}), Eq. (\ref{eq:3compgamalta}) results in
\begin{equation}
\hat{G}(\vect{\Lambda}_{3})\hat{G}(\vect{\Lambda}_{2})\hat{G}(\vect{\Lambda}_{1})=
e^{\alpha_{3}\hat{T}_{+}}e^{\ln(\beta_{3})\hat{T}_{c}}e^{\gamma_{3}\hat{T}_{-}} \, , \\
\label{eq:3compgamalta2}
\end{equation}
where
\begin{align}
\label{beta3}
&\alpha_{3}=\Lambda_{3+}+\frac{\alpha_{2}(\Lambda_{3c})^{\delta}}{1-\epsilon\delta\alpha_{2}\Lambda_{3-}} \, , \:\:\:\:\:\:
\beta_{3}=\frac{\beta_{2}\Lambda_{3c}}{\left(1-\epsilon\delta\alpha_{2}\Lambda_{3-}\right)^{\frac{2}{\delta}}} \,   \nonumber\\[4pt]
&\:\:\:\:\:\:\:\:\:\:\:\:\:\:\:\:\:\:\:\:\:\mbox{and} \:\:\: \gamma_{3}=\gamma_{2}+\frac{\Lambda_{3-}(\beta_{2})^{\delta}}{1-\epsilon\delta\alpha_{2}\Lambda_{3-}} \, .
\end{align}
In order to obtain the result for the composition of $N$ operators, we can proceed by using mathematical induction. After an inspection of Eqs. (\ref{beta2}) and (\ref{beta3}), it is evident that this will be possible. In fact, it is straightforward to show that the composition rule of $N$ $\hat G$ operators, namely, 
\begin{equation}\label{eq:compgroupN}
\hat{G}(\alpha_{N},\beta_{N},\gamma_{N}) = \hat{G}(\vect{\Lambda}_{N})\hat{G}(\vect{\Lambda}_{N-1})\cdots\hat{G}(\vect{\Lambda}_{2})\hat{G}(\vect{\Lambda}_{1})\, ,
\end{equation}
leads to the following recurrence relations
\begin{eqnarray}
\alpha_{N}&=&\Lambda_{N+}+\frac{\alpha_{(N-1)}(\Lambda_{Nc})^{\delta}}{1-\epsilon\delta\alpha_{(N-1)}\Lambda_{N-}} \, , \label{eq:alpharN} \\
\beta_{N}&=&\frac{\beta_{(N-1)}\Lambda_{Nc}}{\left(1-\epsilon\delta\alpha_{(N-1)}\Lambda_{N-}\right)^{\frac{2}{\delta}}} \, , \label{eq:betaN} \\
\gamma_{N}&=&\gamma_{(N-1)}+\frac{\Lambda_{N-}(\beta_{(N-1)})^{\delta}}{1-\epsilon\delta\alpha_{(N-1)}\Lambda_{N-}} \, , \label{eq:gammaN}
\end{eqnarray}
with $\alpha_{1}=\Lambda_{1+}$, $\beta_{1}=\Lambda_{1c}$ and $\gamma_{1}=\Lambda_{1-}$.  
Interestingly, $\alpha$ is an independent term (of $\beta$ and $\gamma$), 
and we can write it in the following form
\begin{equation}
\alpha_{j}=\Lambda_{j+}-\frac{(\Lambda_{jc})^{\delta}}{\epsilon\delta\Lambda_{j-}-\frac{1}{\Lambda_{(j-1)+}-
\frac{(\Lambda_{(j-1)c})^{\delta}}
{\epsilon\delta\Lambda_{(j-1)-} \, -\frac{1}{\ddots \Lambda_{2+}-\frac{(\Lambda_{2c})^{\delta}}{\epsilon\delta\Lambda_{2-}-\frac{1}{\Lambda_{1+}}}}}}}. \\
\label{eq:gammarecursive}
\end{equation}
The above expression is a generalized continued fraction (GCF) for which some topics such as convergence can be investigated. 
GFC's lie in the context of complex analysis and are specially useful to study analyticity of functions as well as number theory among other fields. 
For an interested reader we suggest Ref. \cite{LORENTZEN-1992}. Notice that Eq. (\ref{eq:gammarecursive}) allows an easy numerical implementation, as presented in Ref. \cite{DMT-JPA-2020}.


\section{Application 1: Composition of squeezing operators}

Our first application involves squeezed states, which are of fundamental importance in quantum optics 
(see for instance Refs. \cite{WALLS-1983, DODONOV-2002, Drummond-2004}  and references therein). 
Here we shall use our results to demonstrate the composition law of two arbitrary squeezed operators \cite{Schumaker-1992}. 
Hence, such a demonstration is also a self-consistency check of our main result. 
The demonstration of this important result is rarely found in literature and, as far as the authors' knowledge, it is not 
contemplated in most quantum optics textbooks \cite{Barnett-Book-1997, SCULLY-1997, KLAUDER-2006, LAMBROPOULOS-2007, LEONHARDT-2010, Agarwal-Book-2012}. 

Let us start by introducing briefly the squeezing operators. For an introductory description we suggest Ref. \cite{Barnett-Book-1997}. 
Consider a quantum harmonic oscillator (HO) of unit mass and  frequency $\omega_0$, so that its hamiltonian is given by 
${\hat H}_0 = \frac{1}{2}\left({\hat p}^2 + \omega_0^2{\hat q}^2\right)$ (we are using $\hbar = 1$). 
Using the well known creation and annihilation operators, ${\hat a}^\dagger$ and $\hat a$, which satisfy the commutation relations 
$\left[\hat{a},\hat{a}^{\dagger}\right]=1$ and $\left[\hat{a},\hat{a}\right] = 0 = \left[\hat{a}^\dagger,\hat{a}^{\dagger}\right]$, this hamiltonian can be cast into the form
${\hat H}_0 = \left(\hat n + \frac{1}{2}\right) \omega_0$, where we defined the number operator $\hat n = \hat{a}^{\dagger}\hat{a}$.  
The corresponding energy eigenstates, denoted by $\vert n\rangle$, satisfy the eigenvalue equation 
${\hat H}_0 \vert n\rangle = (n + \frac{1}{2})\omega_0 \vert n\rangle$, with $n=0, 1, ...$.  

A single mode squeezed state $\vert z\rangle$ can be defined by the application of the so-called squeezing operator ${\hat S}(z)$ on the fundamental state, $\vert z\rangle = {\hat S}(z) \vert 0\rangle$, with ${\hat S}(z)$ given by
\begin{equation}
\label{gensqop}
\hat{S}(z)\equiv\exp \left\{-\frac{z}{2}\left.\hat{a}^{\dagger}\right.^{2}+\frac{z^{*}}{2}\hat{a}^{2}\right\} \, ,
\end{equation}
where $z = r e^{i\varphi}$ is a complex number and $*$ stands for complex conjugation. Note that $z$, and hence $r$ and $\varphi$, determines uniquely the squeezed state.  
Using BCH-like relations, it can be shown that the squeezing operator can be written as the following product of exponentials \cite{Barnett-Book-1997} 
\begin{eqnarray}
\label{gensqopbch}
\hat{S}(z)&=&\exp \left\{-e^{i\varphi}\tanh(r) \hat{T}_{+}\right\} \exp \left\{\ln{(\mbox{sech}^{2}(r))} \hat{T}_{c}\right\}\cdot \cr\cr 
&& \cdot \exp \left\{e^{-i\varphi}\tanh(r) \hat{T}_{-} \right\} \, , 
\end{eqnarray}
where we defined
\begin{equation}
\label{algcreann}
\hat{T}_{+} := \frac{\hat{a}^{\dagger^{2}}}{2} \, , \:\:\: \hat{T}_{-} := \frac{\hat{a}^{2}}{2} \:\:\:\:\:\:  \mbox{and}  \:\:\:\:\:\: 
 \hat{T}_{c} := \frac{\hat{a}^{\dagger}\hat{a}+\hat{a}\hat{a}^{\dagger}}{4} \, ,
\end{equation}
which will satisfy, as it can be straightforwardly checked,  the commutation relations written in Eq. (\ref{eq:algebraK}) if we choose  $(\epsilon,\delta)=(1,1)$, so that these operators can be identified as the three generators of the \textit{su}(1,1) Lie algebra. 
Comparing  Eqs. (\ref{eq:BCH1}) and  (\ref{gensqopbch}) results evident that any squeezing operator is a particular case of a 
 $\hat{G}(\vect{\Lambda})$ operator, with parameters $\vect{\Lambda} = (\Lambda_+, \Lambda_c, \Lambda_-)$ given by
\begin{equation}
\label{coeffsque}
\vect{\Lambda} = \left(-e^{i\varphi}\tanh(r),\mbox{sech}^{2}(r),e^{-i\varphi}\tanh(r)\right) \, .
\end{equation}
As a direct consequence of this fact, the composition of $N$ arbitrary squeezing operators is a particular case of a composition of $N$ arbitrary $\hat G$ operators. 
Using Eqs. (\ref{gensqopbch}) and (\ref{coeffsque}), the composition of two squeezing operators, say $\hat S(z_2)\hat S(z_1)$, with $z_1 = r_1 e^{i\varphi_1}$ and $z_2 = r_2 e^{i\varphi_2}$, can be written as
\begin{align}
\hat{S}(z_2)\hat{S}(z_1)\,\,\,=\,\,\,&
e^{\Lambda_{2+}\hat{T}_{+}}e^{\ln(\Lambda_{2c})\hat{T}_{c}}e^{\Lambda_{2-}\hat{T}_{-}} \cdot \nonumber\\[4pt]
&\cdot e^{\Lambda_{1+}\hat{T}_{+}}e^{\ln(\Lambda_{1c})\hat{T}_{c}}e^{\Lambda_{1-}\hat{T}_{-}}  \, ,
\end{align}
where
\small
\begin{equation}
(\Lambda_{j+}\Lambda_{jc}, \Lambda_{j-})
= \left(-e^{i\varphi_j}\tanh(r_j),\mbox{sech}^{2}(r_j),e^{-i\varphi_j}\tanh(r_j)\right) , \nonumber
\end{equation}
\normalsize
with $j=1,2$. Hence, using the new BCH-like relations, Eqs. (\ref{eq:2compgamalta}) and (\ref{beta2}) with $(\epsilon,\delta) = (1,1)$, as well as the last two equations, it is straightforward to show that
\begin{equation}
\label{gopsquee}
\hat{S}(z_{2})\hat{S}(z_{1}) =  e^{\alpha\hat{T}_{+}}e^{\ln(\beta)\hat{T}_{c}}e^{\gamma\hat{T}_{-}}  \, ,
\end{equation}
where
\begin{align}
\label{eqs:coeffcomp}
\alpha=& -\frac{e^{i\varphi_{2}}\tanh(r_{2})+e^{i\varphi_{1}}\tanh(r_{1})}{1+e^{i(\varphi_{1}-\varphi_{2})}\tanh(r_{1})\tanh(r_{2})} \, , \nonumber\\[3pt]
\beta=& \frac{\mbox{sech}^{2}(r_{1})\mbox{sech}^{2}(r_{2})}{\left(1+e^{i(\varphi_{1}-\varphi_{2})}\tanh(r_{1})\tanh(r_{2})\right)^{2}} \, , \nonumber\\[3pt]
\gamma=& \frac{e^{-i\varphi_{2}}\tanh(r_{2})+e^{-i\varphi_{1}}\tanh(r_{1})}{1+e^{i(\varphi_{1}-\varphi_{2})}\tanh(r_{1})\tanh(r_{2})} \, . 
\end{align}
A comment is in order here. For arbitrary $z_1$ and $z_2$, the product of operators $\hat S(z_2)\hat S(z_1)$ written in Eq. (\ref{gopsquee}) 
is not a squeezing operator, except in a very particular case that we shall describe.  This is so because this product of operators cannot 
be cast into the form of Eq. (\ref{gensqopbch}), since parameters  $\alpha$, $\beta$ and $\gamma$ cannot be always written as in Eq. (\ref{coeffsque}). 
In fact, note that Eq. (\ref{coeffsque}) implies the relations 
%
$\left|\alpha\right| = \left|\gamma\right|   \:\:\: \mbox{and} \:\:\: \left|\alpha\right|^{2}+\left|\beta\right| = 1$, 
%
which are not satisfied by the expressions written in Eq. (\ref{eqs:coeffcomp}) for arbitrary values of  $z_1$ and $z_2$. However, it is not difficult to check that if 
$\varphi_1 = \varphi_2 + 2n\pi$ ($n$ integer), then, previous relations hold and in this very particular case, 
we would have $\hat S(z_2)\hat S(z_1) = \hat{S}\left(Re^{i\theta}\right)$, where $R = r_1 + r_2$ and $\theta = \varphi_1 + 2\pi n$, with $n$ an integer.

Coming back to the general case (arbitrary $z_1$ and $z_2$), we shall show that the product $\hat S(z_2)\hat S(z_1)$ 
can be written as a product of a squeezing operator and a rotation operator, 
this last defined by ${\hat R}(\phi) = e^{i\phi {\hat a}^\dagger \hat a}$ ($\phi$ real valued). 
First let us note that a rotation operator can be written as 
\begin{equation}
\hat{R}(\phi) = e^{-i\phi/2}e^{2i\phi \hat{T}_{c}} \, .
\label{eq:RotLie}
\end{equation}
Therefore, except by the phase factor  $e^{-i\phi/2}$, a rotation operator is also a particular case of an operator  $\hat{G}(\vect{\Lambda})$, with coefficients 
\begin{equation}
\vect{\Lambda}=(0,e^{2i\phi},0) \, .
\label{eq:coeffrot}
\end{equation}
Consequently, the product of a squeezing operator and a rotation operator, namely, $\hat{S}(z)\hat{R}(\phi)$, with $z = r e^{i\varphi}$, is a particular case of a composition of two $\hat G$ operators and, hence, this product can be expressed as a $\hat G$ operator multiplied by a phase factor:
\begin{equation}
\label{gopsqueeins}
 \hat{S}(r e^{i\varphi})\hat{R}(\phi) = e^{-i\phi/2} e^{\alpha'\hat{T}_{+}}e^{\ln(\beta')\hat{T}_{c}}e^{\gamma'\hat{T}_{-}} \, ,
\end{equation}
where, after using the new BCH-like relations, Eqs. (\ref{eq:2compgamalta}) and (\ref{beta2}) with $(\epsilon,\delta) = (1,1)$, together with Eqs. (\ref{coeffsque}) and (\ref{eq:coeffrot}), it is straightforward to show that
\begin{eqnarray}
\label{restbetcop}
&&\alpha'= -e^{i\varphi}\tanh(r) \, , \:\:\:\:\:\:
\beta'=  e^{2i\phi}\mbox{sech}^{2}(r) \,   \cr\cr
&&\:\:\:\:\:\:\:\:\:\:\:\:\:\:\: \mbox{and} \:\:\: \gamma'= e^{-i\varphi}\tanh(r) e^{2i\phi} \, .
\end{eqnarray}
Imposing that $(\alpha,\beta,\gamma)=(\alpha',\beta',\gamma')$, \textit{i.e.}, $\hat{S}(z_2)\hat{S}(z_1)=\hat{S}(z)\hat{R}(\phi)$, 
the following relations must be satisfied:
\begin{align}
\label{eqs:phirhodes}
e^{i\varphi}\tanh(r)=&\frac{e^{i\varphi_{2}}\tanh(r_{2})+e^{i\varphi_{1}}\tanh(r_{1})}{1+e^{i(\varphi_{1}-\varphi_{2})}\tanh(r_{1})\tanh(r_{2})} \, , \nonumber\\[3pt]
e^{i\phi}=& \frac{1+e^{i(\varphi_{2}-\varphi_{1})}\tanh(r_{1})\tanh(r_{2})}
{\left|1+e^{i(\varphi_{2}-\varphi_{1})}\tanh(r_{1})\tanh(r_{2})\right|} \, , 
\end{align}
which are precisely the relations quoted in  Eqs. ($3.29$) and ($3.30$) of Ref. \cite{Rhodes-1989}, completing our demonstration. This composition rule is also 
written in Ref. \cite{Agarwal-Book-2012}, as the proposed exercise $3.8$. These results enable us to find $r e^{i\varphi}$ and $e^{i\phi}$, and they are useful, for instance, in analyzing the effect of several down-converters in a cascade arrangement  in the context of SU(1,1) interferometers \cite{Agarwal-Book-2012}. 

We finish this subsection by emphasizing that, although the product of a squeezing operator and a rotation operator can always be written as a $\hat G$ operator, the reciprocal is not necessarily true as it is evident by the previous discussion. 
In fact, it can be shown that an operator $\hat G(\alpha, \beta, \gamma)$ can be written as a product of a squeezing operator and a rotation operator (apart from a phase factor), if coefficients $(\alpha, \beta, \gamma)$ satisfy $\left|\alpha\right|=\left|\gamma\right|  \:\:\: \mbox{and} \:\:\: \beta=\alpha\gamma(1-\left|\alpha\gamma\right|^{-1})$.


\section{Application 2: Time evolution operator of time-dependent quantum systems}\label{Evl}

Let us consider a generic time-dependent hamiltonian that can be written in terms of the generators of the aforementioned Lie algebras
\begin{equation}
\hat{H}(t)=\eta_{+}(t)\hat{T}_{+}+\eta_{c}(t)\hat{T}_{c}+\eta_{-}(t)\hat{T}_{-} \, ,
\label{eq:TDHSU}
\end{equation}
where the $\eta$-coefficients are arbitrary time-dependent complex functions that, together with the specification of the 
algebra, enable us to identify the physical system under consideration. The vector state of the  quantum system satisfies 
the Schr\"odinger equation, namely, $i\frac{\partial}{\partial t}\left|\psi(t)\right\rangle=\hat{H}(t)\left|\psi(t)\right\rangle$ ($\hbar = 1$). 
The corresponding time evolution operator (TEO) is defined as satisfying 
$\left|\psi(t)\right\rangle = \hat{U}(t,0)\left|\psi(0)\right\rangle$. 
Consequently it satisfies the initial condition $ \hat{U}(0,0) = \mathds{1}$. 
It is immediate to see that  $\hat{U}(t,0)$ obeys  the  differential equation, 
$i\frac{\partial}{\partial t}\hat{U}(t,0)=\hat{H}(t)\hat{U}(t,0)$, 
whose formal solution is given by the so called Schwinger-Dyson equation, 
$\hat{U}(t,0) = {\cal T} \exp\left\{-i\int_0^t \hat H(t^\prime) dt^\prime\right\}$, 
where ${\cal T}$ means time ordering operator. Looking at the Schwinger-Dyson equation 
one realizes that it is not an easy task to solve the Schr\"odinger equation 
for time-dependent hamiltonians. However, there are other approaches. 
Particularly, in the following discussion we shall show that, whenever the hamiltonian can be written as in Eq. (\ref{eq:TDHSU}), 
the corresponding TEO results to be a $\hat{G}$ operator. Therefore, using the results of Section II, 
one can obtain iteratively the TEO and, consequently, the dynamics of the system 
at any instant of time and with the desired precision.


Initially, notice that the TEO satisfies the composition property which follows directly from its definition,  namely,
\begin{equation}
\hat{U}(t,0)=\hat{U}(t,t_{N-1})\hat{U}(t_{N-1},t_{N-2})\cdots\hat{U}(t_{1},0)\, .
\label{eq:compoTEO}
\end{equation}
The above expression is exact, no matter the number of time intervals. For finite time intervals the expressions for  $\hat{U}(t_j,t_{j-1})$, with $j = 1,2,...,N-1$, could be very complicated, since the problem under consideration has a time-dependent hamiltonian. However,  if we take 
$\tau\rightarrow 0$ and $N\rightarrow\infty$, with $N\tau = t$, we can write the TEO as an infinite product of simple infinitesimal TEO's, namely, 
\begin{equation}
\hat{U}(t,0)  = \lim_{\substack{N\rightarrow\infty\\ N\tau = t}}e^{-i\hat{H}(N\tau)\tau}e^{-i\hat{H}\left((N-1)\tau\right)\tau}\cdots\;e^{-i\hat{H}(\tau)\tau} \, . \nonumber\\
\label{eq:TEOin}
\end{equation}
%

Let us write the time-dependent coefficients of Eq. (\ref{eq:TDHSU}) in the following compact way
\begin{equation}
\vect{\eta}(t)=(\eta_{+}(t),\eta_{c}(t),\eta_{-}(t)) \, .
\label{eq:complamb}
\end{equation}
Now, let us consider $N$ large enough so that in each time interval $\tau = t/N$ (without any loss of generality we are considering equally sized intervals)  the time-dependent 
coefficients $\vect{\eta}(t)$ can be assumed as fixed in time, namely,
\begin{eqnarray}
\vect{\eta}(t)=\left\{
\begin{array}{ccc}
\vect{\eta}_{0}\, , \: & \mbox{for} & t\leq 0, \\
\vect{\eta}_{1}\, , \: & \mbox{for} & 0<t\leq\tau, \\
 \vdots & & \vdots \\
\vect{\eta}_{j}\, , \: &\mbox{for} & (j-1)\tau<t\leq j\tau, \\
 \vdots & & \vdots \\
\vect{\eta}_{N}\, , \: & \mbox{for} & (N-1)\tau<t\leq N\tau \, , \\
\end{array} \right.
\label{eq:discrete}
\end{eqnarray}
where $\vect{\eta}_j=(\eta_{j+},\eta_{jc},\eta_{j-})$ can 
be chosen as any value of $\vect{\eta}(t)$ within the interval $t_{j-1} < t \le t_j$. For convenience, we choose the extreme value $\vect{\eta}_{j} :=\vect{\eta}(j\tau)$. 
Recall that $N\tau=t$, and the exact result is obtained only in the limit  $N\rightarrow\infty$ (and $\tau\rightarrow = 0$).  
Once $t_j - t_{j-1} = \tau$, for any $j$ we may approximate the hamiltonian in each time interval $t_{j-1} < t \le t_j$, denoted by $H_j$,  as a constant one. Hence, from Eq. (\ref{eq:TDHSU}) we may write
\begin{equation}
\hat{H}_{j}=\eta_{j+}\hat{T}_{+}+\eta_{jc}\hat{T}_{c}+\eta_{j-}\hat{T}_{-} \, .
\label{eq:TDHSUdiscre}
\end{equation}
Since all $\hat{H}_j$ are now considered as time-independent hamiltonians, the TEO for each time interval ${\hat U}(t_j,t_{j-1})$, with $j = 1,2,...,N$, can be written as
\begin{equation}\label{eq:TEOj}
\hat{U}_{j} := {\hat U}(t_j,t_{j-1}) = e^{-i\hat{H}_{j}\tau} = e^{\lambda_{j+}\hat{T}_{+}+\lambda_{jc}\hat{T}_{c}+\lambda_{j-}\hat{T}_{-}}.
\end{equation}
Notice that the above expression can be identified as a $\hat{G}(\vect{\lambda})$ operator, 
in the representation of Eq. (\ref{eq:Gj}), with $\vect{\lambda}_j=-i\tau\vect{\eta}_j$. 
Using the above equation in Eq. (\ref{eq:compoTEO}), the TEO at an arbitrary instant $t>0$ is given by 
\begin{align}
\hat{U}(t,0) \,\,\,=\,\,\, & \hat{U}_{N}\hat{U}_{(N-1)}\cdots\hat{U}_{2}\hat{U}_{1} \nonumber\\[3pt]
=\,\,\, & e^{\left\{\lambda_{N+}\hat{T}_{+}+\lambda_{Nc}\hat{T}_{c}+\lambda_{N-}\hat{T}_{-}\right\}}\cdot \nonumber\\[3pt]
& \cdot e^{\left\{\lambda_{(N-1)+}\hat{T}_{+}+\lambda_{(N-1)c}\hat{T}_{c}+\lambda_{(N-1)-}\hat{T}_{-}\right\}}\;  \cdots \nonumber\\[3pt]
&\cdots\;  e^{\left\{\lambda_{2+}\hat{T}_{+}+\lambda_{2c}\hat{T}_{c}+\lambda_{2-}\hat{T}_{-}\right\}}\cdot \nonumber\\[3pt]
&\cdot e^{\left\{\lambda_{1+}\hat{T}_{+}+\lambda_{1c}\hat{T}_{c}+\lambda_{1-}\hat{T}_{-}\right\}} \, . 
\label{eq:TEOComp1}
\end{align}
Remembering from Eqs. (\ref{eq:Gj}) and (\ref{eq:BCH1}) that 
\begin{equation}
e^{\lambda_{+}\hat{T}_{+}+\lambda_{c}\hat{T}_{c}+\lambda_{-}\hat{T}_{-}} = 
e^{\Lambda_{+}\hat{T}_{+}}e^{\ln(\Lambda_{c})\hat{T}_{c}}e^{\Lambda_{-}\hat{T}_{-}}  \, ,
\label{f=h}
\end{equation}
where the $\lambda$'s and $\Lambda$'s are related by Eqs. (\ref{truej4})-(\ref{truej6}), Eq. (\ref{eq:TEOComp1}) can be cast into the form
\begin{align}
\hat{U}(t,0)\,\,\, =\,\,\, & e^{\Lambda_{N+}\hat{T}_{+}}e^{\ln(\Lambda_{Nc})\hat{T}_{c}}e^{\Lambda_{N-}\hat{T}_{-}}\cdot \nonumber\\[3pt]
&\cdot  e^{\Lambda_{(N-1)+}\hat{T}_{+}}e^{\ln(\Lambda_{(N-1)c})\hat{T}_{c}}e^{\Lambda_{(N-1)-}\hat{T}_{-}}\;  \cdots \nonumber\\[3pt]
&\cdots\;  e^{\Lambda_{2+}\hat{T}_{+}}e^{\ln(\Lambda_{2c})\hat{T}_{c}}e^{\Lambda_{2-}\hat{T}_{-}} \cdot \nonumber\\[3pt]
&\cdot e^{\Lambda_{1+}\hat{T}_{+}}e^{\ln(\Lambda_{1c})\hat{T}_{c}}e^{\Lambda_{1-}\hat{T}_{-}} \, .
\label{eq:finalTEO}
\end{align}
Since each line on the right hand side of the previous equation can be identified as a $\hat{G}(\vect{\Lambda})$ operator in the representation of Eq. (\ref{eq:BCH1}), 
we can use the relations developed before in Sec. II for the composition of N $\hat{G}$ operators, to write this equation into the form
\begin{equation}
\hat{U}(t,0) = e^{\alpha_{N}\hat{T}_{+}}e^{\ln(\beta_{N})\hat{T}_{c}}e^{\gamma_{N}\hat{T}_{-}} \, ,
\label{eq:TEONcomp}
\end{equation}
where the coefficients $\alpha_{N}$, $\beta_{N}$ and $\gamma_{N}$ are given by Eqs. (\ref{eq:alpharN}), (\ref{eq:betaN}) and (\ref{eq:gammaN}). 
Since these equations are in fact a system of recurrence relations, the previous expression for the TEO can be computed iteratively up to the desired precision. 
This iterative way of solving such complex time-dependent problems can be easily implemented numerically (see Ref. \cite{DMT-JPA-2020} for explicit examples of a particular case of this formula).

\section{CONCLUSIONS} \label{C}

Using ordering techniques we deduced new BCH-like relations arising from the composition of two elements of the 
\textit{SU}$(1,1)$, \textit{SU}$(2)$ and \textit{SO}$(2,1)$ Lie groups. 
In order to make a unified demonstration for these three cases, 
we introduced two parameters whose different values characterize the different algebras, 
in the same spirit as in Truax's paper \cite{Truax-1985}.
Then, we used our results to obtain, in a straightforward way, a rule to calculate the composition 
of an arbitrary number of elements. 
This rule is given as an iterative analytical solution and we showed that it can be written 
in terms of a general continued fraction.
A particular case of great interest in quantum optics, namely, the arbitrary composition of squeezing operators, 
is contemplated by our general result. Furthermore, as a self-consistency check of our main result, 
we presented an original way to obtain the composition law of two squeezing operators. 
Finally, we have developed a new recipe that enables to compute iteratively the corresponding TEO of a quantum system described 
by a time-dependent hamiltonian that can be written in terms of the generators of the aforementioned Lie algebras. 
We hope our results provides, not only an alternative and convenient way for solving the considered physical systems, 
but also paves the way for the construction of similar methods for other time-dependent hamiltonians, \textit{i.e}, 
for other Lie algebras.


\begin{acknowledgments}

The authors acknowledge L. Pires and M. V. Cougo-Pinto  for enlightening discussions. 
The authors thank the Brazilian agency for scientific and technological research CAPES for partial financial support.
This work was partially supported by Conselho Nacional de Desenvolvimento Cient\'{\i}fico e Tecnol\'{o}gico - CNPq, 310365/2018-0 (C.F.) and 309982/2018-9 (C.A.D.Z.)

\end{acknowledgments}


\end{document}